# A Framework for Agricultural Food Supply Chain using Blockchain

Sudarssan N
UG Student, Department of Artificial
Intelligence and Data Science
Sri Eshwar College of Engineering
Coimbatore, India
sudarssan73@gmail.com

*Abstract*— The main aim of the paper is to create a trust and transparency in the food supply chain system, ensuring food safety for everyone with the help of Blockchain Technology. Food supply chain is the process of tracing a crop from the farmer or producer to the buyer. With the advent of blockchain, providing a safe and fraud-free environment for the provision of numerous agricultural necessities has become much easier. Because of the globalization of trade, the present supply chain market today includes various companies involving integration of data, complex transactions and distribution. Information tamper resistance, supply-demand relationships, and traceable oversight are all difficulties that arise as a result of this. Blockchain is a distributed ledger technology that can provide information that is resistant to tampering. This strategy can eliminate the need for a centralized trusted authority, intermediaries, and business histories, allowing for increased production and security while maintaining the highest levels of integrity, liability, and safety. In order to have an integrity and transparency in food supply chain in the agricultural sector, a framework is proposed here based on block chain and IoT.

## I. INTRODUCTION

Agriculture is one of the most essential areas in the world that needs to be prioritized in order for the majority of mankind to exist. Agriculture output has a significant impact on a country's economy, as well as its citizens' security, nutrition, and health. Weather varies from season to season, market prices for farm goods fluctuate, soil quality deteriorates, crops are not sustainable, weeds and pests impair yield, and global climate change are all factors that affect agriculture practice [1].

For the agriculture sector to boost production and sustainability, it is becoming increasingly important to leverage data and information. In agriculture, information and communication technologies (ICT) such as Blockchain, Internet of Things (IoT), Big data, Wireless Sensor Networks (WSNs), Cloud Computing, and Machine Learning significantly improve the effectiveness and efficiency of data collection, storage, analysis, and use [2]. All these technologies can help to improve the agricultural food supply chain and quality management [3].

With the advent of blockchain, food systems can regain their confidence and transparency. It allows for information traceability in the food supply chain, which helps to improve food safety. It promotes the creation and deployment of data-driven technologies for smart farming and smart index-based agriculture insurance by providing a secure manner of storing and managing data. Big data analytics could be used in the agriculture supply chain to monitor food quality, storage conditions, weather patterns in a specific geographic area, soil quality (such as pH and nutrients), and marketing. In order to increase the quality and productivity of farming, WSNs are frequently utilized in agriculture monitoring. Sensors collect various types of data such as humidity, $CO_2$ level, and temperature to analyze in real time.

IoT is intended to assist farmers in analyzing and planning more efficient irrigation, as well as making harvest forecasts using data obtained from remote sensors such as humidity, air temperature, and soil quality. Cloud computing can be used to aggregate, analyze and store data from soil sensors, satellite images, and weather stations to help farmers make better decisions about managing their crops. Cloud computing in IoT facilitates seamless communication between IoT devices. This enables many robust APIs to interact between connected devices and smart devices. Cloud-connected wireless sensors capture data from the field and machine learning algorithms analyze that real-time information, giving farmers a better understanding of crops conditions. It also allows farmers to leverage the internet to reduce waste, better pest control, streamline livestock management, and increase productivity.

Agricultural practitioners and farmers can use ICT to access up-to-date information and make better judgments in their daily operations. The remote data collected related with the condition of soil can help farmers manage their crops [4]. Farmers' access to markets and financial support is improved because to mobile phones, which minimize the cost of information [5]. Despite the fact that blockchain technology has acquired popularity as a result of its function in the financial sector, it offers a wide range of applications outside of crypto currencies. Many industries, including healthcare, law, real estate, banking, and others, are expected to be dramatically transformed by technology. The agriculture sector can benefit from blockchain technology in a variety of ways [6]. Distributed ledgers and smart contracts, which are part of blockchain technology in agriculture, have the potential to weed out counterfeits in agri-food production and supply chains, resulting in healthier products for consumers, generating trust between business players, and enabling a better life on a global scale [7].

Crop and food production, food supply chain management, weather crisis management, and agricultural finance management are all possible applications of blockchain in agriculture. The focus is on the food supply chain in this instance. Supply chains connect businesses to their suppliers, allowing them to manufacture and deliver goods and services. A supply chain is a set of actions that lead to the delivery of a product or service to a consumer. Consumer and government worries about food quality have gain interest in the concept of supply-chain traceability. However, due to its intrinsic qualities of decentralization, transparency, and immutability, blockchain plays a crucial role in the growth of supply chains.

It also offers smart contracts that make use of safe trading transactions between entities. Despite the lack of confidence in blockchain based Agri Food supply chains, end-consumers find it difficult to trust the product owner and quality of the product before making a purchase. As a result of this survey paper, a framework is modeled here to have an agriculture trade in the absence of third parties by providing a transparent distribution system.

The remainder of the paragraphs is arranged as follows, section II deals with related works in the block chain based supply chain system, third section deals with the concept of blockchain, its benefits for agriculture and framework of block chain. Fourth section concentrates on the proposed framework and the fifth is the conclusion.

## II. RELATED WORKS

In this segment, the research related with the usage of blockchain for food supply chain management is discussed. Agriculture today is becoming digital technology. As it turns out, the most severe challenges bring the greatest opportunities for innovation in agriculture. Dabbene and Gay [8] claim that accurate data gathering via information communication methods like bar-codes and RFID allows for enhanced data acquisition and traceability in agricultural and food supply chains. The existing practice of traceability in the agricultural supply chain is hampered by data fragmentation and centralized controls, making it susceptible to data manipulation and management. Individual stages in food supply chains are frequently traceable, but information sharing between stages is complex and time-consuming [9]. Recent technological advancements, such as the use of blockchain technology, can provide a significant and practical solution for maintaining agricultural output traceability while also removing the requirement for a trusted centralized authority [10].

The authors in [11] utilized smart contracts to govern and control all transactions with respect to soybean supply chain among all the participants. A detailed survey regarding the security for IoT based smart farming based on blockchain is given in [12]. It featured the applications of IoT in agriculture and blockchain evolution in smart agriculture. The work given in [13] elaborated the trust, transparency in the food supply chain and the issues related with the security.

A new model based on blockchain for agriculture supply chain is discussed in [14] for making the smart contract as an efficient one. The challenges in adopting the blockchain for the agriculture is discussed in [15, 16, 17] and also they came up with a optimized solution for it. Decreasing third-party engagement in the supply chain system and increasing data security was the main motive behind the research which is detailed in [18]. It was stated that there are a number of serious faults in traditional supply chain management methods.

The disadvantages in the traditional systems were listed as lack of traceability, trouble maintaining product safety and quality, failure to monitor and control inventory in warehouses and stores, escalating supply chain expenses, and so on. The study presented in above was mainly to reduce third-party involvement and to improve the data security in the supply chain system. Throughout the process, this enhances accessibility, efficiency, and timeliness. The main findings from the literature can be said as if a transparent trading is possible, and then the people will feel more secure during the payment procedure in the food supply chain in agriculture.

## III. ABOUT BLOCKCHAIN

The crypto currency Bitcoin propelled Blockchain technology to popularity in early 2009. To move money, Bitcoin users with a changeable Public Key (PK) generate and broadcast transactions to the network. Users have pushed these transactions into a block. When a block is full, a mining procedure is used to append the block to the Blockchain. To mine a block, a group of nodes called miners attempt to solve a resource-intensive cryptographic puzzle known as Proof of Work. When a new transaction occurs, the Blockchain technology allows all members to preserve a ledger including all transaction data and to update their ledgers to ensure integrity.

The single point of failure resulting from reliance on an approved third party has been eliminated thanks to the advancement of the Internet and encryption technology, which allows all members to check the reliability of a transaction. Because many individuals own the transaction information, hacking is more difficult. Security costs are reduced, transactions are automatically approved and documented by public involvement, and promptness is ensured. Furthermore, the system is simple to set up, connect, and grow thanks to open source and transaction records.

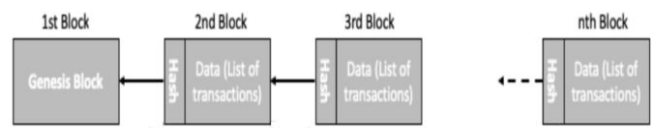

Fig. 1. Block functionality based on hash value

It is extremely difficult to falsify and alter the registered data since the hash values saved in each peer in the block are affected by the values of prior blocks. Figure. 1 depicts the block functionality based on the hash value. The hash value is generated for every block based on the content and it will be referred in the subsequent blocks. The block header will have the details as shown in Fig. 2. The generation and processing of a block, as well as the blockchain is detailed in [19]. It also explained how the hash of the previous block is linked to the hash of the subsequent block. Although data change is conceivable if 51% of peers are hacked simultaneously, the attack scenario is realistically quite challenging [20].

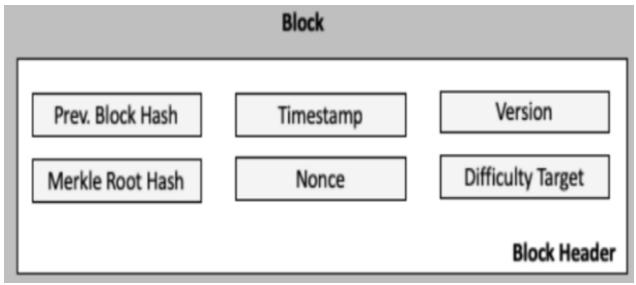

Fig. 2. Block Header Details

### A. Benefits of Blockchain for Agriculture

The blockchain technology provides a secure way to track transactions between anonymous parties. As a result, fraud and defects can be identified promptly. Furthermore, smart contracts can be used to report problems in real time [21, 22]. Due to the complexity of the agri-food system, this helps to overcome the difficulty of tracing products across a large supply chain. As a result, the technology addresses concerns of food quality and safety, which are of great importance to consumers, governments, and others. Diagrammatic representation of the benefits of blockchain for agriculture is shown in Figure. 3.

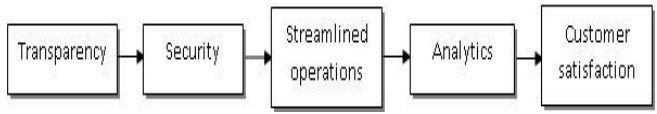

Fig. 3. Benefits of Blockchain for Agriculture

The blockchain technology promotes the collection of trustworthy data and provides transparency to all parties concerned. From the point of creation to the point of death, blockchain can track every stage of a product's value chain. The reliable data of the farming process are highly valuable for developing data-driven facilities and insurance solutions for making farming smarter and less vulnerable.

### B. Limitations of Block Chain

The blockchain technology allows for information traceability in the food supply chain, which helps to improve food safety. It promotes the creation and deployment of data-driven technologies for smart farming and smart index-based agriculture insurance by providing a secure manner of storing and managing data.

Furthermore, it has the potential to lower transaction costs, improving farmers' access to markets and providing new revenue sources. Despite its immense potential benefits, blockchain technology in agriculture and food still has significant limits. First, further research is needed into the motivations of the transacting parties to contribute accurate and real data to the blockchain ledger. This is particularly relevant in the case of small-scale farming.

### C. Blockchain Frameworks

Blockchain frameworks are a type of software that makes it easier to design, implement, and support technically sophisticated products. The blockchain framework is the foundation upon which any blockchain application can be built; it provides the feature of transparent and immutable record distribution. Everyone who has time-stepped is linked to the previous one. Members of the marketplace can use the features to check the exchanges without having to leave the platform.

The following are the Blockchain Frameworks which can be chosen for the decentralized applications. They are Ethereum, Bigchain DB, Corda, Hyperledger Fabric, Hyperledger Sawtooth, Quorum. The selection of blockchain framework can be tricky — and a deep analysis has to be done before making a choice. The factors that have to be taken care are 1. Community, 2. License, 3. Support model, 4. Roadmap, 5. Activity and 6. Ease of use.

## IV. PROPOSED BLOCK CHAIN FRAMEWORK FOR SUPPLY CHAIN

Blockchain in agriculture can be realized in different stages of farming such as (i) crop and food production, (ii) food supply chain, (iii) controlling weather changes and (iv) managing the financial issues in agriculture. The main objective of the paper here is to model a framework for food supply chain. Blockchain is used here to upgrade the current trade, and to make all the transactions transparent. As a result, it can be said that definitely, it can bring faith among people in the trade.

The following are the steps in the process of blockchain based food supply:

- Data generated by IoT devices or data stored by farmers
- Gown crop distribution to food processing firms
- Processed food distribution to wholesalers and retailers
- The supply chain can be traced by consumers.

**Step 1:** Data generated by IoT devices or data stored by farmers

Initially, the sensors collect the real time data about the crops in the agricultural field. The data is related with crop quality, variety of seed and the weather conditions when the crops were sown. These collected data will be stored in the blockchain.

**Step 2:** Gown crop distribution to food processing firms

After harvesting or during harvesting season, food processing firms will start their bidding process. The crops will be transferred to the concern firm through the IoT enabled vehicles. Once the bidding process is over, and it is checked through smart contracts and the companies will store all these data in the block chain.

**Step 3:** Processed food distribution to wholesalers and retailers

By checking with the firms, the wholesalers or retailers will verify whether the food is of good quality or not. The data processing starting from the initial stage to the processing stage is stored in blockchain. Hence, the data can be checked at any time they need.

**Step 4:** The supply chain can be traced by consumers

All these processes are digital in nature, hence consumers can visualize everything; because all the stages of food supply chain process is stored in blockchain. Consumers can find the details of the farm, vehicles used for the transportation, nature of food processing, temperature at which the foods are processed and stored, expiry details.

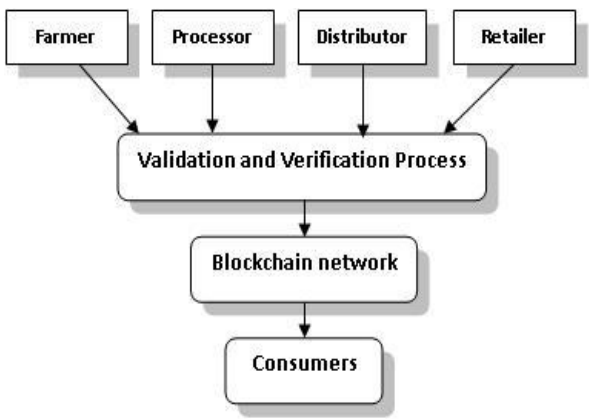

Fig. 4. Process Flow in Blockchain based Food Supply Chain

Figure 4 shows the process flow in a blockchain based food supply chain. All the transactions are authenticated and it is transparent to the consumers.

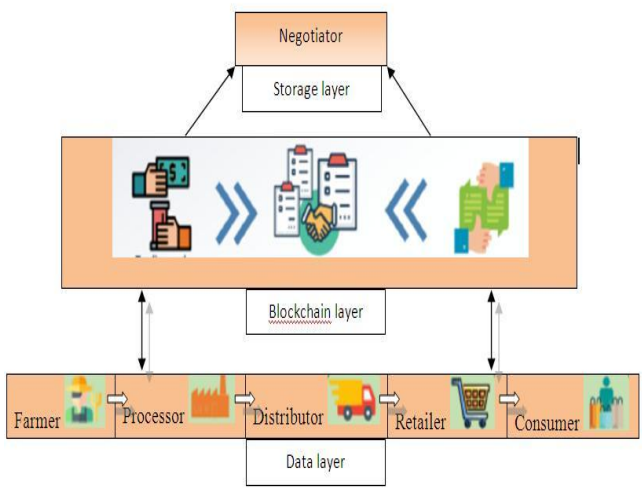

Fig. 5. Framework model for Agricultural Food Supply Chain based on Blockchain

Fig. 5 is a framework for the blockchain based food supply chain system. It consists of data layer, blockchain layer and storage layer. Data layer consists of farmer, processor, distributor, retailer and consumer. Blockchain layer consists of trading and delivery system, smart contracts, reputation system. The storage layer contains only negotiator who will act as a dispute handler.

## V. CONCLUSION

The expected outcome from the proposed framework model is to simplify the current complexity in agricultural goods transportation, taking into account differences in stakeholders' requirements in rural and urban transit, and to reduce highly technological infrastructure demands through efficient use of mobile phones and additional functionalities such as GPS to share information among stakeholders. At last to achieve a common language and standard information exchange patterns, and to reduce the transportation and intermediary costs associated with routing by sharing agro-goods information that is easily deployable and immediately available.